\documentclass[prd,
showpacs
]{revtex4}
\usepackage{graphicx}

\def\hbar{\hspace{0pt}\raisebox{1pt}{$-$} \hspace{-7pt} h}

\def\5{\overline 5}

\newcommand{\be}{\begin{equation}}
\newcommand{\ee}{\end{equation}}
\newcommand{\ba}{\begin{eqnarray}}
\newcommand{\ea}{\end{eqnarray}}
\newcommand{\no}{\nonumber}

\begin{document}
\title[ On the stability of thick brane worlds non-minimally coupled to gravity]{On the stability of thick brane worlds non-minimally coupled to gravity}
\date{\today
}
\author{A. A. Andrianov\footnote{On leave of absence from V.A. Fock Department of Theoretical Physics, St. Petersburg State University,
 Russia;\\ andrianov@bo.infn.it}}
\affiliation{Departament d'Estructura i Constituents de la Mat\`eria and\\
 Institut de Ci\`encies del Cosmos, Universitat de Barcelona,\\
 Diagonal 647, 08028 Barcelona, Spain}
\author{L.Vecchi}
\affiliation{INFN, Sezione di Trieste and\\
Scuola Internazionale Superiore di Studi Avanzati\\
via Beirut 4, I-34014 Trieste, Italy}
\begin{abstract}
We analyze a class of 5D models where a 3 brane is generated by a bulk scalar field non minimally coupled to gravity. We show that perturbative stability of such  branes is normally guaranteed although non minimal couplings are not innocuous in general. After the physical states are identified the linearized equations for propagating modes are evaluated into a Schr\"odinger form and supersymmetric quantum mechanics provides the absence of tachyons. The spectrum contains a tower of spin 2 and spin 0 fields with continuous masses starting from zero ones. For regular geometries the scalar spectrum contains a state with zero mass which is always non normalizable. The propagating massive scalar states are repelled off the brane due to a centrifugal potential.
\noindent
\end{abstract}
\preprint{SISSA ??/??/EP}
\maketitle
%
\large
\section{Introduction and motivations}

Brane worlds \cite{rubsh}--\cite{RS} open a way to understand a number of long-standing problems in particle physics such as
fermion mass hierarchy and the smallness of cosmological constant (see reviews \cite{rev1}--\cite{rev10}).
The branes themselves  may exhibit their specific excitations - branons which contribute into the discovery potential of high-energy colliders \cite{bran3}.

Theoretically brane worlds could well be created spontaneously if the bulk in extra dimensions is filled not only by gravity but also by primordial \cite{gremm1}-\cite{scal3} or composite \cite{aags} scalar matter self-interacting so that its condensation  breaks the translational invariance (in the form of a kink for one extra dimension). Such a configuration is essential also to trigger localization of fermions \cite{rubsh} and possibly of other matter fields \cite{dewolfe,scal3,bounce,galoc3}.  In the absence of gravity the latter localization holds perfectly also for scalar fields in the Goldstone boson sector related to spontaneous breaking of  translational invariance . When
matter induced gravity affects the geometry in the bulk, the scalar Goldstone mode mix strongly with scalar components of multi-dimensional gravity and can be removed by a gauge  choice \cite{Giovannini}. The physical scalar zero-mode
fluctuation apparently disappears from the particle phenomenology.

The latter mechanism has been analyzed in spontaneously generated brane worlds with minimal gravitational interaction. However if both gravity and scalar fields $\Phi$ are induced by more fundamental
matter fields then at low energies, from vacuum polarization effects, one recovers also a non-minimal interaction between space-time curvature $R$
and scalar fields $\xi R \Phi^2$ . This is a purpose of our work to
examine the particle spectrum in interplay of gravity and a bulk scalar field, in the case of the above non-minimal interaction, when a brane world is generated spontaneously. We analyze the perturbative stability against quantum fluctuations, i.e. the absence of tachyons in the spectrum, as well as
the phenomenon of (de)localization of light particles on a brane.

One can find that such a term is not in general innocuous. We show that in the case of constant scalar configurations such a term generally causes instability of the scalar field. Contrary to the previous case, a non minimally coupled scalar matter may be perturbatively stable if its vacuum configuration is non trivial. We prove this in the particular case of brane world scenarios.

It seems to be possible to use the well established perturbative stability of the
$\xi=0$ system \cite{Giovannini},\cite{Freedman} and make a conformal transformation to eliminate
the $\xi$ term from the action. Doing this, however, we introduce
non analytical interactions for the scalar field. Since the theorem
for the perturbative equivalence of the on-shell S-matrix is
rigorously proved only in the case of analytic change of field
variables, we believe that the most trustable way to face the
problem is by showing explicitly the positivity of the spectrum.

The paper is organized as follows.

In section II we introduce our notation and the equations of motion (EOM) for general scalar and gravitational backgrounds. We briefly discuss the effect of the non minimal coupling for the case of constant scalar configurations and introduce the ansatz for the brane configuration.

In section III the linear gauge invariance is analyzed. The quadratic action in the oscillations around the background is constructed by invoking the gauge symmetry. As its consequence not all linear equations derived are independent.

In section IV the decoupling procedure for the quadratic action is performed. We show that the gravitational field can be decoupled from the scalar oscillations after a field redefinition. Two different gauges are then chosen to simplify the analysis of the decoupled system, one of which is very convenient to study the spectrum.

In section V we discuss the spectrum of the brane scenario. This is composed by a tower of massive spin 2 fields (gravitons) and a tower of spin 0 fields (branons), as expected. We show that both spectra are determined completely by a single function of the warped factor and of the scalar v.e.v. The sign of this function determines the positivity of the kinetic and mass terms for both  subsystems. The linearized equations are converted into a Schr\"odinger form and the localization of wavefunctions is discussed.

\section{Notations and EOM}

We assume the 5D action to have the form: \be I[g,\Phi]=\int d^5X\
{\cal L}(g,\Phi),\label{act} \ee with \ba {\cal L} &=& \sqrt{\vert
g\vert}\left\lbrace
-M_{\ast}^3R+\partial_A\Phi\partial^A\Phi-V(\Phi)+\xi R\Phi^2
\right\rbrace. \label{lagrangian}\ea A cosmological constant can be considered as englobed into the scalar potential $V$.

The 5D coordinates are denoted by $X^A=(x^\mu,z)$, with Greek indices $\mu,\nu,\dots=0,1,2,3$ and Latin indices $A,B,\dots=0,1,2,3,5$.

The equations of motion (EOM) are: \ba
R_{AB}-\frac{1}{2}g_{AB}R=\frac{1}{M_\ast^3}T_{AB}\\ \no
D^2\Phi=-\frac{1}{2}\frac{\partial V}{\partial\Phi}+\xi R\Phi,\ea
where $D^2=D_CD^C$ and $D_C$ is a covariant derivative. The
energy momentum tensor reads \ba
T_{AB}&=&\partial_A\Phi\partial_B\Phi-\frac{1}{2}g_{AB}\left(\partial_C\Phi\partial^C\Phi-V(\Phi)\right)\\
\no
&+&\xi\left(R_{AB}-\frac{1}{2}g_{AB}R+g_{AB}D^CD_C-D_AD_B\right)\Phi^2 .\ea

The presence of a non minimal coupling governed by the parameter $\xi$ may cause  destabilization of scalar configurations \cite{farakos}. This is a general
characteristic of constant scalar backgrounds and does not depend
on the specific geometry of space-time. To see this we make the
trace of the Einstein equation to find the scalar curvature.
Substituting it in the EOM for the scalar field we can find
it in full generality as: \ba D^2\Phi=-U'_{eff}\ea with \ba
U'_{eff}=\frac{\frac{1}{2}(-M^3_\ast+\xi\Phi^2)V'+\xi\Phi\frac{d}{2-d}V+
\xi\Phi\left(1+\xi\frac{4(d-1)}{2-d}\right)
\Phi_{,A}\Phi^{,A}}{-M^3_\ast+\xi\Phi^2\left(1+\xi\frac{4(d-1)}{2-d}\right)},
\ea with $d$ being the dimensionality of space-time and $V'$ denoting the derivative with respect to the scalar field. The effective
potential $U_{eff}$ determines the stability of a scalar configuration
although it does not correspond to its physical energy.

The existence of a global minimum for $U_{eff}$ requires this
function to grow ($U'_{eff}\geq0$) in the limit
$\Phi\rightarrow+\infty$. For a constant solution \ba
U'_{eff}\rightarrow\frac{\frac{1}{2}\Phi
V'+\frac{d}{2-d}V}{\Phi\left(1+\xi\frac{4(d-1)}{2-d}\right)}.\ea
If we assume $V\propto
\lambda\Phi^n$ with $\lambda>0$ and $n>2d/(d-2)$ then the numerator is always a positive quantity. A necessary
condition for having a global minimum is therefore
$1+\xi\frac{4(d-1)}{2-d}>0$. For $d=5$ we get the condition
$1-\xi\frac{16}{3}>0$.

In the case of non trivial scalar configurations the above method cannot be straightforwardly applied since the stability is controlled by a functional.

Let's limit ourselves to the study of background solutions which don't
spontaneously break the 4D Poincar{\`e} invariance and take
$g_{AB}=A^2(z)\eta_{AB}$, with $\eta_{AB}=diag(1,-1,-1,-1,-1)$,
and $\Phi=\Phi(z)$. The equations of motion in terms of this
ansatz read \be \frac{1}{2}A^5\left(\frac{\Phi'^2}{A^2}-V(\Phi)
\right)+6AA'^2\left(-M^3_\ast+\xi\Phi^2\right)+4\xi A^2A'(\Phi^2)'
=0, \label{EOMS}\ee \be
-\frac{1}{2}A^5\left(\frac{\Phi'^2}{A^2}+V(\Phi)
\right)+3A^2A''\left(-M^3_\ast+\xi\Phi^2\right)+2\xi
A^2A'(\Phi^2)'+\xi A^3(\Phi^2)''=0,\label{EOMh}\ee \be
2\left(A^3\Phi'\right)'-A^5\frac{\delta V}{\delta\Phi}+\left(16\xi
A^2A''+8\xi AA'^2\right)\Phi=0 \label{EOMPhi}\ee where from now on $f'=df/dz$.

Since we have three equations for two unknown functions, one of the
above conditions must be redundant. This is a consequence of gauge invariance and can be seen explicitly by
subtracting (\ref{EOMh}) from (\ref{EOMS}) to derive an expression for
the potential. Differentiating it one can recover (\ref{EOMPhi}).

We select out the solutions with a definite parity in the fifth direction in respect to $z=0$, i.e. the potential $V(\Phi)$ is ${\textbf{Z}}_2$ symmetric.

\section{Quadratic action}

By construction, the action~(\ref{act}) is invariant under general
diffeomorphisms. Because the space-time variable $X$ is a dummy
variable the symmetry can be seen as an invariance of the action
under appropriate transformations of the fields.
This transformation is the Lie derivative along an arbitrary vector $\zeta^A$ defined by the coordinate transformation $X\rightarrow \tilde X=X+\zeta(X)$. To the first order one finds:
\ba
\tilde g_{AB}(X)&=&
g_{AB}(X)- \zeta_{,A}^C\ {g}_{CB}(X)-\zeta_{,B}^C\ {g}_{AC}(X)
- {g}_{AB,C}(X)\ \zeta^C+O(\zeta^2)\\ \no &=& g_{AB}(X) -
\zeta_{A;B}-\zeta_{B;A}+O(\zeta^2)
\ea
where '$;$' denotes the covariant derivative.

Let's consider the general case of non
trivial backgrounds $\bar{g}_{AB}(X)$ and define the fluctuating field $h_{AB}(X)$ as follows:
\ba g_{AB}(X) \equiv \bar{g}_{AB}(X) + h_{AB}(X). \ea The Lie derivative acting on $h_{AB}(X)$ is highly non linear and maybe expanded in powers of $\zeta,h$. Since the action of our interest is up to quadratic order in the fluctuations, we confine ourselves to the leading order: \ba \tilde h_{AB}(X)= h_{AB}(X) -
\zeta_{A;B}-\zeta_{B;A}+O(\zeta^2,h\zeta)\label{gauge} \ea where now the '$;$' denotes the covariant derivative
with respect to the background $\bar{g}_{AB}(X)$.

The same line of reasoning applies to general tensors. We take the fluctuations around the solution of the EOM to
be: \be g_{AB}(X)=A^2(z)\left( \eta_{AB}+h_{AB}(X)\right) ;\quad
\Phi(X)=\Phi(z)+\phi(X). \ee Since the 4D symmetry is unbroken we
adopt the 4D notation $h_{5\mu}\equiv v_\mu$, $h_{55}\equiv S$.
Introducing the notation $\hat\zeta_A$, with
$\zeta_\mu=A^2\hat\zeta_\mu$ and $\zeta_5=A\hat\zeta_5$ for convenience we can express our gauge symmetry as follows \ba h_{\mu\nu}&\rightarrow&
h_{\mu\nu}-\left(\hat\zeta_{\mu,\nu}+\hat\zeta_{\nu,\mu}-\frac{2A'}{A^2}\eta_{\mu\nu}\
\hat\zeta_5 \right)\label{gaugetr} \\ \no v_\mu&\rightarrow& v_\mu
-\left(\frac{1}{A}\hat\zeta_{5,\mu}+\hat\zeta'_\mu \right)\\ \no
S&\rightarrow& S - \frac{2}{A}\hat\zeta'_5\\ \no
\phi&\rightarrow&\phi+\hat\zeta_5\frac{\Phi'}{A}.  \ea
Notice that these transformations are exact up to $O(\zeta^2,h^2,h\zeta)$ terms (where $h$ stands for an arbitrary fluctuation). The symmetry transformations leaving the solution $h_{AB}=\phi=0$ unchanged are the isometries of the background metric. It is easy to see that they are restricted to the 4D Poincare group. This fact justifies our decomposition of the fields under $SO(1,3)$ representations.

After gauging, the unbroken 4D Poincare group implies the existence of a 4D massless spin 2 field which has to be identified as a graviton. From the above gauge transformations we expect this state to be described by  $z$-independent fluctuations of $h_{\mu\nu}$. This is intuitively understood since the 4D space is flat and this field actually describes the space-time dependent fluctuations of $\eta_{\mu\nu}$.

Since the translations along $z$ are spontaneously broken a Goldstone boson (GB) must appear. We expect only one scalar state because the Lorentz rotations orthogonal to $z$, though spontaneously broken, don't act independently on the vacuum and therefore don't generate additional massless states~\cite{Man}. As usual, the GB can be identified as the space-time dependent coordinate $\zeta_5(x)=A\hat\zeta_5(x,z)$, where the 4D dependence ensures that its propagation is confined to the unbroken directions. However, due to the gauge nature of the symmetry, the GB is locally gauge equivalent to the zero solution,therefore it is not a physical state and must be rotated away.

We will show in the next section that once the gauge symmetry is completely fixed (the GB has been absorbed) the propagating fields will describe a tower of spin 2 and spin 0 particles.

\subsection{Quadratic action}

We now expand the action in Taylor series up to quadratic order
in field fluctuations. As already discussed, its gauge invariance at leading order
in $\zeta$ also depends on the higher order terms in the fluctuations. At the second order in  fluctuations we can get rid of the latter ones by simply
imposing the equations of motion.

The transformations (\ref{gaugetr}) can help to eliminate gauge degrees
of freedom \cite{Boos} .
To simplify  the direct calculations as much as possible we adopt the gauge $S=v_\mu=0$ and explicitly calculate the
quadratic lagrangian~(\ref{lagrangian}) using the formulas in Appendix A (see also \cite{Callin}). For a further
convenience we write it as a sum of two pieces: \ba {\cal
L}_{(2)}(v_\mu=S=0)={\cal L}_{h}+{\cal L}_{\phi},\ea where \ba {\cal
L}_{h}&\equiv&\ A^3\left(-M^3_\ast+\xi\Phi^2 \right)\ \left\lbrace
-\frac{1}{4}\ h_{\alpha\beta,\nu}h^{\alpha\beta,\nu} -\frac{1}{2}\
h^{\alpha\beta}_{,\beta}h_{,\alpha}+\frac{1}{2}\
h^{\alpha\nu}_{,\alpha}h^{\beta}_{\nu,\beta}+\frac{1}{4}\
h_{,\alpha}h^{,\alpha}\right\rbrace\label{h}\\ \no &+& \ A^3\left(-M^3_\ast
+\xi\Phi^2\right)\ \left\lbrace
\frac{1}{4}h'_{\mu\nu}h'^{\mu\nu}-\frac{1}{4}h'^2\right\rbrace\ea
and \ba  {\cal L}_{\phi}&\equiv& A^3 \phi_{,\mu}\phi^{,\mu}-A^3 \phi'^2
- \frac{1}{2}\ A^5
\frac{\delta^2 V}{\delta\Phi^2}(\Phi) \phi^2 \label{phi} \\
\no &+& \left(A^3\Phi' +8\xi A^2A'\Phi \right)h'\phi + 2\xi
A^3\Phi\left(h^{\mu\nu}_{,\mu\nu}-h^{,\mu}_{,\mu}+h''\right)\phi.
\ea The commas denote partial derivatives and all indices are
raised up with the Minkowski flat metric. In particular, $h=h_{\mu\nu}\eta^{\mu\nu}$.

We can construct the quadratic action in an arbitrary gauge by
invoking gauge invariance and using the iterative procedure we now present. The linear terms in the vector field
can be evaluated as follows. After a gauge transformation defined
by $\hat{\zeta}_\mu(x,z)$ the fields $h_{\mu\nu}$ and $v_\mu$ are
changed in such a way that: \ba \delta I=\int\left\lbrace
\frac{\delta{ I}}{\delta
h_{\mu\nu}}\left(-\hat\zeta_{\mu,\nu}-\hat\zeta_{\nu,\mu}
\right)+\frac{\delta{ I}}{\delta
v_{\mu}}\left(\hat\zeta'_{\mu}\right)+O(\hat\zeta^2,hh\hat\zeta)\right\rbrace=0.
\ea It is easy to see that this condition is satisfied for any
$\hat\zeta_\mu$ only if: \ba \frac{\delta{I}}{\delta
v^\mu}(v_\mu=S=0)&=&A^3(
-M^3_\ast+\xi\Phi^2)\left(h_{\mu\nu}^{,\nu}-h_{,\mu} \right)'\\
\no &+& \left(2A^3\Phi'+16\xi
A^2A'\Phi\right)\phi_{,\mu}-4\xi(A^3\Phi\phi_{,\mu})'. \ea

An analogous procedure is applicable when we perform a gauge transformation
defined by $\hat\zeta_5(x,z)$. In this case  the
variation with respect to $S$ can be derived if we take into account the
transformation of $\phi$, too. One gets, \ba \frac{\delta{I}}{\delta
S}(v_\mu=S=0)&=&-\frac{1}{2}\ A^3
\left(-M^3_\ast+\xi\Phi^2\right)\left(h^{\mu\nu}_{,\mu\nu}-h^{,\mu}_{,\mu}
\right)-\frac{1}{2}\left( A^3\left(-M^3_\ast+\xi\Phi^2 \right)\right)'h'\\
\no &+& (A^3\Phi'\phi)' - 2A^3\Phi'\phi'+2\xi
A^3\Phi\phi_{,\mu}^{,\mu}-8\xi (A^2A'\Phi\phi)'\\
\no &+&(16\xi A^2A''+8\xi AA'^2)\Phi\phi.\label{S} \ea
The quadratic part in $S$ is obtained by requiring that
the linear terms in $S$ automatically cancel under a gauge
transformation. This is satisfied if and only if the full
quadratic action for the $S$ field in the $v_\mu=0$ gauge is, \be
{\cal L}_{S}\equiv -\frac{1}{4}A^5VS^2+ \frac{\delta{I}}{\delta
S}(v_\mu=S=0)S.\ee
With the inclusion of ${\cal L}_{S}$ to the quadratic action, the
first derivative in $v_\mu$ receives additional contributions.
Repeating the procedure outlined we find, \ba \frac{\delta{I}}{\delta
v^\mu}(v_\mu=0)&=&A^3(
-M^3_\ast+\xi\Phi^2)\left(h_{\mu\nu}^{,\nu}-h_{,\mu} \right)'\\
\no &+& \left(2A^3\Phi'+16\xi
A^2A'\Phi\right)\phi_{,\mu}-4\xi(A^3\Phi\phi_{,\mu})'\\\no
&-&\left[A^3(-M^3_\ast+\xi\Phi^2)\right]'S_{,\mu}. \label{V} \ea
From this latter we derive the quadratic action in $v_\mu$: \ba {\cal L}_{V}\equiv \frac{1}{4}A^3\left(-M^3_\ast+\xi\Phi^2\right)v_{\mu\nu}v^{\mu\nu}+ \frac{\delta{I}}{{\delta
v}^\mu}(v_\mu=0)v^\mu,\ea where $v_{\mu\nu}=v_{\mu,\nu}-v_{\nu,\mu}$.

The full action to the  quadratic order represents finally the sum of
(\ref{h}), (\ref{phi}), (\ref{S}) and (\ref{V}), \ba {\cal L}_{(2)}= {\cal
L}_{h}+{\cal L}_{\phi}+{\cal L}_S+{\cal L}_{V}\label{fullL}.\ea This result can be explicitly checked exploiting the formulas in Appendix A, but we stress that its form is completely fixed by gauge invariance. As
we have seen, gauge invariance implies that the linearized equations are not all independent.
This point will be very useful when solving the coupled mass
eigenvalue equations. It is also important to emphasize that the gauge conditions can be imposed already at the lagrangian level if the secondary constraints are also taken into account.

\section{Decoupling}

The physical spectrum can be identified after the system being completely decoupled. It can be achieved  by a redefinition of the
gravitational field. We will see that the particle spectrum of these models comprise a tower of spin 0 and spin 2 fields.

To unravel completely the
quadratic action one has to shift $h_{\mu\nu}$ by a special
solution of its equation of motion, \ba
&&\frac{1}{2}A^3\left(-M^3_\ast+\xi\Phi^2 \right) \left\lbrace \
h_{\alpha\beta,\mu}^{,\mu} -\ h^{,\mu}_{\alpha\mu,\beta}-\
h^{,\mu}_{\beta\mu,\alpha}+ h_{,\alpha\beta}+\ \eta_{\alpha\beta}
h^{\mu\nu}_{,\mu\nu}-\
\eta_{\alpha\beta}h^{,\mu}_{,\mu}\right\rbrace\label{LEh}\\ \no &&-
\frac{1}{2}\left[A^3\left(-M^3_\ast +\xi\Phi^2\right) \left(
h'_{\alpha\beta}-\eta_{\alpha\beta}h'\right)\right]'+\frac{1}{2}\left[A^3\left(-M^3_\ast +\xi\Phi^2\right) \left(
v_{\alpha,\beta}+v_{\beta,\alpha}-2\eta_{\alpha\beta}v_\mu^{,\mu}\right)\right]'\\\no
&&-\eta_{\alpha\beta}\left[A^3\Phi'\phi+8\xi
A^2A'\Phi\phi\right]'+\ 2\xi
A^3\Phi(\phi_{,\alpha\beta}-\eta_{\alpha\beta}\phi_{,\mu}^{,\mu})+2\xi\eta_{\alpha\beta}(A^3\Phi\phi)''\\\no&&-\frac{1}{2}A^3\left(-M^3_\ast+\xi\Phi^2
\right) \left(S_{,\alpha\beta}-\eta_{\alpha\beta}S^{,\mu}_{,\mu}
\right)+\frac{1}{2}\eta_{\alpha\beta}\left\lbrace\left[A^3\left(-M^3_\ast+\xi\Phi^2
\right)\right]'S\right\rbrace' =0. \ea
This solution can be written in terms of two scalars $E$, $\psi$ and a vector $F_\mu$,
\ba h_{\mu\nu}\rightarrow
h_{\mu\nu}+F_{\mu,\nu}+F_{\nu,\mu}+E_{,\mu\nu}+\eta_{\mu\nu}\psi\ea
for $F'_\mu=v_\mu$ and:
\ba \Xi\psi-\frac{1}{2}\Xi S-\frac{1}{2}(\Xi E')'+ 2\xi A^3\Phi\phi =0\label{cond}\\
\no 3\Xi\psi'+\Xi'S=2A^3\Phi'\phi+16\xi
A^2A'\Phi\phi-4\xi(A^3\Phi\phi)',\ea
where the convenient notation $\Xi=A^3\left(-M^3_\ast+\xi\Phi^2 \right)$ has been introduced. The first(second) condition follows from the off diagonal (diagonal) terms of equation~(\ref{LEh}).

Substituting the redefined tensor field in~(\ref{fullL}) one can rewrite the lagrangian as the sum of a tensorial contribution (${\cal L}_{grav}$) and an action containing the scalar fields $S,\phi,E,\psi$ and $v_\mu$. Because of the above constraints ~(\ref{cond}) only two scalars out of four are independent and, making use of the gauge freedom defined by $\hat\zeta_5$, we conclude that only one of them actually describes a propagating degree of freedom. The vector $v_\mu$ is a $\hat\zeta_\mu$-gauge variable and can be rotated away.

After the decoupling conditions are imposed we can choose an arbitrary gauge to simplify the analysis of the spectrum. In the light-cone gauge the result should agree with~\cite{STZ}. Here we decide to work with $v_\mu=0$. This choice leaves a residual freedom parametrized by $\hat\zeta_5$ and $\hat\zeta_\mu$ with $A\hat\zeta'_\mu+\hat\zeta_{5,\mu}=0$, see~(\ref{gaugetr}).

Now let's examine two simple $\hat\zeta_5$-gauge choices. The first one is defined by setting $S$ to be an appropriate function of $\phi$ which implies $\psi=0$. In this case the physical branon is given by a $\phi$ field and the analysis of the constant $\Phi$ solutions turns out to be easy. To study the non trivial v.e.v. case (brane solutions) the most convenient choice is $\phi=0$. In this case the branon field is described by $\psi$.

\subsection{General scalar background in $\psi=0$ gauge}

Setting $\psi=0$ we can interpret the second equation~(\ref{cond}) as a gauge choice on $S$.
The conditions~(\ref{cond}) define $S$ and $E$ as
non local functions of $\phi$. The physical branon turns out to be described by $\phi$.

By inserting our shifted tensor field in the quadratic action the
linear terms in the physical $h_{\mu\nu}$ are canceled by
construction. Moreover, all contributions containing the field $E$
automatically cancel because of the above conditions (\ref{cond}).
The quadratic action can therefore be written as the sum of the
graviton contribution plus a scalar part describing the physical
branon: \ba {\cal L}_{grav}&=&\ A^3\left(-M^3_\ast+\xi\Phi^2
\right)\ \left\lbrace -\frac{1}{4}\
h_{\alpha\beta,\nu}h^{\alpha\beta,\nu} -\frac{1}{2}\
h^{\alpha\beta}_{,\beta}h_{,\alpha}+\frac{1}{2}\
h^{\alpha\nu}_{,\alpha}h^{\beta}_{\nu,\beta}+\frac{1}{4}\
h_{,\alpha}h^{,\alpha}\right\rbrace\\ \no &+& \
A^3\left(-M^3_\ast+\xi\Phi^2 \right)\ \left\lbrace
\frac{1}{4}h'_{\mu\nu}h'^{\mu\nu}-\frac{1}{4}h'^2\right\rbrace\ea
and \ba {\cal L}_{bran}&=& A^3 \phi_{,\mu}\phi^{,\mu}-A^3 \phi'^2
- \frac{1}{2}\ A^5 \frac{\delta^2 V}{\delta\Phi^2}(\Phi) \phi^2
+(8\xi A^2A''+4\xi AA'^2)\phi^2-\frac{1}{4}A^5 V(\Phi)S^2 \\\no
&+&\left\lbrace(A^3\Phi'\phi)' - 2A^3\Phi'\phi'+2\xi
A^3\Phi\phi_{,\mu}^{,\mu}-8\xi (A^2A'\Phi\phi)'+(16\xi A^2A''+8\xi
AA'^2)\Phi\phi\right\rbrace S. \ea
The residual gauge invariance defined by $\hat\zeta'_\mu=0$ will be fixed in the next section.

The gravitational field is now completely decoupled from the
scalar degrees of freedom. Expressing $S$ in terms of the field
$\phi$ and integrating by parts we get: \ba {\cal L}_{bran}={\tilde
A}^3 \left\lbrace\phi_{,\mu}\phi^{,\mu}-\phi'^2 -{\tilde
U}\phi^2\right\rbrace . \ea The explicit expressions are rather
complicated and read: \ba {\tilde A}^3=
A^3-\frac{2\xi}{\Xi'^2}\left\lbrace\left[A^6(\Phi^2)'+16\xi
A^5A'\Phi^2\right]\Xi'-2\xi A^6\Phi^2\Xi''\right\rbrace\ea and
\ba
{\tilde A}^3{\tilde U}&=&A^3\left[\frac{1}{2}A^2V''-8\xi H'-12\xi
H^2\right]\\\no &-& \frac{A^6}{\Xi'}\left[(1-2\xi)\Phi'+2\xi
H\Phi\right]\left\lbrace
A^2V'-\frac{A^5V}{\Xi'}\left[(1-2\xi)\Phi'+2\xi H\Phi\right]-24\xi
H^2\Phi-16\xi H\Phi'\right\rbrace \\\no &+& \left\lbrace
2\xi\frac{A^{11}V}{\Xi'^2}\Phi\left[(1-2\xi)\Phi'+2\xi
H\Phi\right]-\xi\frac{A^8V'}{\Xi'}\Phi\right.\\\no&& \left.-
\frac{A^6}{\Xi'}\left[(1-2\xi)\Phi'^2-8\xi^2
H^2\Phi^2+(10\xi-32\xi^2)H\Phi\Phi'\right]\right\rbrace'.\ea In
the above definitions the expression $V'$ denotes the variation
with respect to the field $\Phi$. The quantities $H=A'/A$ and
$\Xi=A^3\left(-M^3_\ast+\xi\Phi^2 \right)$ have been used for
brevity and the relation $\Xi''=A^5V$ has also been employed.

Rescaling the field $\phi=\tilde A^{-3/2}\Psi$ we obtain the
standard form: \ba {\cal
L}_{bran}&=&\Psi_{,\mu}\Psi^{,\mu}-\Psi'^2 -U\Psi^2\ea with \ba
U=\frac{3}{4}\frac{\tilde A'^2}{\tilde
A^2}+\frac{3}{2}\frac{\tilde A''}{\tilde A}+\tilde U.\label{potU}\ea

The stability of the configuration is not manifest and will be
analyzed later on. For the moment we study two simple limits of
the potential: the case $\xi=0$ and $\Phi'=0$.

For $\xi=0$ we have $\hat A^3\rightarrow A^3$,
$\Xi\rightarrow-M_\ast^3A^3$ and \ba {\tilde A}^3{\tilde
U}\rightarrow
A^3\left\lbrace\frac{1}{2}A^2V''+\frac{1}{3HM_\ast^3}\left[A^2V'\Phi'+\frac{1}{3}\frac{A^2V}{HM_\ast^3}\Phi'^2\right]+
\frac{1}{3A^3}\left[\frac{A^3\Phi'^2}{HM_\ast^3}\right]'\right\rbrace.\ea
Making use of the EOM we get: \ba
U\rightarrow\frac{\Omega''}{\Omega},\quad
\Omega=\frac{A^{3/2}\Phi'}{2H}.\ea This result agrees with~\cite{Giovannini}.
In the next section we will derive the generalization of this
expression for the case of arbitrary $\xi$.

Although physically less interesting, the case $\Phi'=0$ reveals
some general feature. We make use of the approximation
$\xi\Phi^2\ll M_\ast^3$ to simplify the result. It is then
straightforward to evaluate the potential: \ba
U\rightarrow\frac{15}{4}H^2\left(1-\frac{16}{3}\xi\right)+\frac{1}{2}A^2V''+O\left(\xi\frac{\Phi^2}{M^3_\ast}\right).\ea
A sufficient condition for stability is that $\Phi$ be a local
minimum of the potential $V$ and $\xi<3/16$. For $\xi$
sufficiently large we see that the flat geometry $A'=0$ is
favored. This represents the local version of the result
obtained in section II.

\subsection{Non-trivial scalar background in $\phi=0$ gauge}

We now turn to the study of the physical spectrum.
A simple check in the EOM reveals two possible solutions for the scalar v.e.v.: a trivial one and a non trivial one. The latter condition in particular requires that $\Phi'=0$ only in isolated points. The gauge $\phi=0$ can be satisfied almost everywhere. We now show that the propagating graviton field is described by $\psi$.

By inserting our shifted tensor field and setting $\phi=0$, the
action simplifies considerably and can be written as the sum of
the graviton contribution plus a scalar part: \ba {\cal
L}_{(2)}(v_\mu=\phi=0)={\cal L}_{h}+ {\cal L}_{S}= {\cal L}_{grav}+
{\cal L}_{bran}\ea where \ba {\cal L}_{grav}&=&\Xi \left\lbrace
-\frac{1}{4}\ h_{\alpha\beta,\nu}h^{\alpha\beta,\nu} -\frac{1}{2}\
h^{\alpha\beta}_{,\beta}h_{,\alpha}+\frac{1}{2}\
h^{\alpha\nu}_{,\alpha}h^{\beta}_{\nu,\beta}+\frac{1}{4}\
h_{,\alpha}h^{,\alpha}\right\rbrace\\ \no &+& \Xi \left\lbrace
\frac{1}{4}h'_{\mu\nu}h'^{\mu\nu}-\frac{1}{4}h'^2\right\rbrace\ea
and \ba {\cal L}_{bran}&=&
\frac{3}{2}\Xi\psi_{,\mu}\psi^{,\mu}-3\Xi\psi'^2+\frac{3}{2}\Xi\psi_{,\mu}^{,\mu}S-2\Xi'\psi'S-\frac{1}{4}\Xi''S^2.
\ea The first two terms in the latter expression come from the
quadratic part in the tensor field (${\cal L}_{h}$) while the last
three from the terms linear in $S$ (${\cal L}_{S}$). The identity
$\Xi''=A^5V$ has also been used.

Expressing $S$ in terms of the
derivative of the field $\psi$, exploiting (\ref{cond}) and
integrating by parts in the 4D variables we get: \ba {\cal
L}_{bran}={\hat \Omega}^2
\left\lbrace\psi_{,\mu}\psi^{,\mu}-\psi'^2\right\rbrace  \ea where
\ba {\hat \Omega}^2=
-3\Xi+\frac{9}{4}\frac{\Xi''\Xi^2}{\Xi'^2}=A^3\left(\frac{3}{2}\frac{\Xi\Phi'}{\Xi'}\right)^2-3\Xi\left(\xi
A^3\frac{(\Phi^2)'}{\Xi'}\right)^2.\ea  In the last equality the EOM
have been employed.

By studying the linearized equations in the $\phi=0$ gauge one can actually find a decoupled condition for $S$ and realize that its solution is a divergent function. In our notation its divergence can be traced back to the relation $S\sim \Xi\psi'/\Xi'$ following from~(\ref{cond}). Since the physical state $\psi$ is delta function normalizable, $S$ will diverge as $1/H\sim z$ for asymptotically constant scalar configurations. The field $S$ cannot be projected out, as it was assumed in~\cite{Kak}, because it is not an independent configuration. Therefore the conclusion of~\cite{Kak} on the necessity of having conformal matter on the brane is not well justified.

Introducing the rescaled field $\psi= {\hat \Omega}^{-1}\Psi$ we
obtain: \ba {\cal L}_{bran}= \Psi_{,\mu}\Psi^{,\mu}- \Psi'^2
-U\Psi^2\ea where \ba U= \frac{\hat \Omega''}{\hat \Omega}\ea is forced by gauge
invariance to coincide with~(\ref{potU}). It is easy to verify it in the case $\xi=0$ ($\hat\Omega\rightarrow\Omega$)
while for the case $\Phi'=0$ is not trivial.

\subsection{Note}

In order to better understand the above results we briefly discuss the degrees of freedom involved.
It is convenient to decompose the 15 gravitational fields in terms of its
traceless-transverse tensor, vectors and scalar components: \ba
h_{\mu\nu}=b_{\mu\nu}+f_{\mu,\nu}+f_{\nu,\mu}+E_{,\mu\nu}+\eta_{\mu\nu}\psi\ea
where $b_{\mu\nu}$ and $f_\mu$ satisfy
$b_{\mu\nu}^{,\mu}=b=0=f_{\mu}^{,\mu}$. The $h_{5A}$ fields are
still denoted as $v_\mu$ and $S$.

Substituting this form in the full quadratic action one can recognize $f_\mu$ and $E$ as auxiliary fields. $E$ is in fact a
Lagrange multiplier and gives rise to a constraint which is the second equation of~(\ref{cond}). Gauge
invariance requires this latter to be equivalent to the condition
$\frac{\delta{I}}{\delta v^\mu}(v_\nu=0)=0$, indicating that the
graviphoton appears in the quadratic action only via its kinetic
term. The vector $f_\mu$ turns out not to be coupled to any
field. Its integration does not lead to interesting
relations and we will simply ignore it in this discussion.

The above mentioned constraint is of extreme importance because it
involves the three scalars $\psi, S$ and $\phi$ implying that only
two of them are independent. Since the field $S$ contains no
kinetic term while $\psi$ does, it may be natural to choose $\psi$
and $\phi$ as the independent variables. The initial 15 + 1 scalar
degrees of freedom are now reduced to a traceless and transverse
field $b_{\mu\nu}$, a vector $v_\mu$ and two scalars $\psi, \phi$.

The physical degrees of freedom can be identified after the gauge
is completely fixed. For a non compact extra dimension the vector field $v_\mu$ can be completely gauged away by an appropriate choice of $\hat\zeta_\mu$. The residual gauge symmetry depends on $\hat\zeta_5$ and acts on the independent scalar components as \ba
\psi&\rightarrow&\psi+ \frac{2A'}{A^2} \hat\zeta_5\\ \no
\phi&\rightarrow&\phi+\hat\zeta_5\frac{\Phi'}{A}. \ea A gauge dependent
combination of $\psi$ and $\phi$ can then be eaten and the orthogonal combination can be chosen to be the gauge invariant field \ba \Psi\propto
\phi-\frac{A\Phi'}{2A'}\psi.\ea This is the natural candidate to
describe the physical branon~\cite{Giovannini}. In the previous subsections we have shown the explicit form of the quadratic action for this field.

\section{Eigenstates}

\subsection{Spin 2 fields}

$\Xi$ determines the sign of the quadratic Hamiltonian for both
branons and gravitons. In order to have a positive definite
quadratic energy we require $\Xi<0$. The opposite sign may
indicate a breaking of our semiclassical analysis and it would
necessitate a quantum gravity justification. We therefore assume
from here on that $\xi\Phi^2\ll M_\ast^3$.

Introducing the rescaled field $h_{\mu\nu}= (-\Xi)^{-1/2}\sqrt{2}b_{\mu\nu}$
we obtain:\ba {\cal L}_{grav}&=&\frac{1}{2}\
b_{\alpha\beta,\nu}b^{\alpha\beta,\nu} +
b^{\alpha\beta}_{,\beta}b_{,\alpha}-
b^{\alpha\nu}_{,\alpha}b^{\beta}_{\nu,\beta}-\frac{1}{2}\
b_{,\alpha}b^{,\alpha}\\ \no &-&\left\lbrace
\frac{1}{2}\left(b'_{\mu\nu}b'^{\mu\nu}+Wb_{\mu\nu}b^{\mu\nu}\right)-\frac{1}{2}\left(b'^2+Wb^2\right)\right\rbrace\ea
with \ba W=
\frac{1}{2}\frac{\Xi''}{\Xi}-\frac{1}{4}\frac{\Xi'^2}{\Xi^2}=K^2-K', \quad
K=-\frac{1}{2}\frac{\Xi'}{\Xi}.\ea The action can be put in
standard form if we define \ba b_{\mu\nu}(X)=\sum_m
b_{\mu\nu}^{(m)}(x)b_m(z) \ea where the wavefunctions $b_m(z)$
satisfy the eigenvalue equation \ba \label{WFGrav}&&
-b_m''+Wb_m=(-\partial_z+K)(\partial_z+K)b_m=m^2b_m\\ \no &&\int
dz b_mb_{m'}=\delta_{m,m'},\ea
with manifestly positive masses.

The zero mode is:
\ba
b_0(z)= C_{(1)}\Xi(z)^{1/2}+C_{(2)}\Xi(z)^{1/2}\int^z dz'\frac{1}{\Xi(z')}
\ea
and it is a physical state if its wave function is normalizable. In the semi-classical approximation $\xi\Phi^2\ll M_\ast^3$ we can analyze the convergence by studying the regularity and the asymptotic behavior of $A^3$. It is easy to see that the integral~(\ref{WFGrav}) can be convergent only if one of the two constants $C_{(1,2)}$ is zero.

In the above basis the 5D bulk dynamics can be integrated leaving an
effective 4D action describing a tower of spin 2 massive states
whose quadratic action reads: \ba I_{grav}&=&\sum_m\int d^4x
\left\lbrace\left( \frac{1}{2}\
{b^{(m)}}_{\alpha\beta,\nu}{b_{(m)}}^{\alpha\beta,\nu} +\
{b_{(m)}}^{\alpha\beta}_{,\beta}{b^{(m)}}_{,\alpha}-\
{b_{(m)}}^{\alpha\nu}_{,\alpha}{b_{(m)}}^{\beta}_{\nu,\beta}-\frac{1}{2}\
{b^{(m)}}_{,\alpha}{b_{(m)}}^{,\alpha}\right) \right.\\ \no &&-
\left. \frac{m^2}{2}\left(
{b^{(m)}}_{\mu\nu}{b_{(m)}}^{\mu\nu}-{b_{(m)}}^2\right)
\right\rbrace.\label{Lbb} \ea As already discussed, the
propagating fields are transverse-traceless as appropriate for
massive spin 2 states. The
remnant gauge freedom acting on it as $b^0_{\mu\nu}\rightarrow
b^0_{\mu\nu}-\hat\zeta_{\mu,\nu}-\hat\zeta_{\nu,\mu}$ with
$\zeta'_\mu=0$ represents the usual gauge symmetry of the 4D
graviton field. It is defined by a transverse $\hat\zeta_\mu(x)$ satisfying the free scalar
field EOM. This is exactly what is needed in order to further reduce
the d.o.f. of the on-shell graviton by three units.

\subsection{Spin 0 fields}

The
lagrangian describing the spin 0 field is: \ba {\cal L}_{bran}=
\Psi_{,\mu}\Psi^{,\mu}- \Psi'^2 -U\Psi^2\ea where \ba U=
\frac{\hat \Omega''}{\hat \Omega}=J^2-J', \quad J=-\frac{\hat \Omega'}{\hat
\Omega}.\ea This result explicitly shows the positivity of the scalar
spectrum.

We expand $\Psi$ in an appropriate basis: \ba
\Psi(X)=\sum_m \psi^{(m)}(x)\Psi_m(z)\ea with \ba &&
-\Psi_m''+U\Psi_m=(-\partial_z+J)(\partial_z+J)\Psi_m=m^2\Psi_m\\
\no && \int dz \Psi_m\Psi_{m'}=\delta_{m,m'}.\ea

The lightest state is the zero mode:
\be\label{zeromode}
\Psi_0(z)= D_{(1)}{\hat \Omega}+D_{(2)}{\hat \Omega}\int^z dz'\frac{1}{{\hat \Omega}^{2}}.
\ee
Again, the integration constants $D_{(1)},D_{(2)}$ cannot be both different from zero. One can easily see from the EOM that the asymptotic form of $\hat \Omega$ is the same as that of $A^{3/2}$. We conclude that once the graviton solution has been chosen ($C_{(1)}=0$ or $C_{(2)}=0$) the scalar zero mode with a well defined limit at infinity is unique ($D_{(1)}=0$ or $D_{(2)}=0$). But, in fact, the two solutions $D_{1,2}$ cannot satisfy simultaneously the convergence at zero
and infinity, each of them can satisfy one of the requirements if a smooth limit for $A(z)$ is required. To see this we follow~\cite{GiovanniniLoc} and expand our solution near $z=0$:
\ba A=1+az^\alpha+\cdots\\ \no \Phi=b+cz^\beta+\cdots\ea
where $\beta>0$ guarantees a smooth $z=0$ limit for the graviton wavefunction. Imposing the EOM we find $\alpha=2\beta$ and $\hat \Omega^2\sim1/z^\alpha$. The mode~(\ref{zeromode}) is singular at the origin and must be projected out from the dynamics of the model. The absence of the scalar zero mode was first noticed in~\cite{Giovannini} (see also \cite{Kim}).

In order to determine the existence of a mass gap we have to study the asymptotics of $A^3\sim\hat \Omega^2$. We again follow~\cite{GiovanniniLoc} and notice that, for any power law $A\sim z^{-\gamma}$ with $\gamma>0$, the scalar potential is $U\sim1/z^2$ and the spectrum turns out to be a continuous starting at zero. For an exponential behavior of $A$ we may have a mass gap. A reasoning to fix the asymptotics of the warp factor may be the requirement of having non singular curvature invariants at any point in the fifth dimension~\cite{GiovanniniLoc}. Assuming this, one has to rule out exponentially varying $A(z)$ and confine the study to power laws. The spectrum which follows is continuous and ranges from zero to infinity.

The massive states behave as a linear combination of regular and irregular Bessel's functions at infinity.
Near the origin the potential effectively acts as a repulsive centrifugal force $U\sim\omega^2/z^2$ expelling the wavefunctions of the massive spectrum. They behave like regular Bessel's functions on the brane. This delocalization effect is valid for any value of the 5D Planck scale, namely $U$ does not depend on $M_\ast$.

\section{Conclusions}

We have shown that the brane solutions generated by a non trivial scalar
configuration remain stable after the inclusion of the non-minimal $\xi
R\Phi^2$ coupling, for any $\xi$. This is contrary to what happens
for trivial scalar backgrounds, where a general $\xi$ determines
unboundedness of the effective potential felt by the scalar
field.

The quadratic action for the model has been constructed by imposing gauge invariance.
The non trivial scalar v.e.v. together with the non minimal coupling generate a mixing between the fundamental scalar and the scalar components of the metric. A judicious choice
of gauge can simplify the decoupling and the analysis of the spectrum. For non-trivial backgrounds only, we can find a reference frame in which the excitation of the fundamental scalar is
absent and where the role of the physical branon is taken by a conformal scalar component of the 4D excitations of the metric.
The 55 component is never a physical state.

The spectrum contains a tower of massive spin 2 (gravitons) and spin 0 (branons) fields. The normalizability and absence of a mass gap for the
fields depends on the assumption of asymptotic AdS (plus regularity in the case
of the branon). The graviton spectrum is always continuous and possesses a normalizable zero mode. For regular geometries the scalar excitations also form a continuous ranging from zero to infinity but the zero mode is never normalizable and must be projected out. The physical massive scalar states are repelled off the origin and cannot be relevant for the localized 4D particle dynamics.

\acknowledgments
We are grateful to V.A. Andrianov, P. Giacconi and R. Soldati for useful and inspiring discussions at the early stage of our work. 
This work is supported (L.V.) by MIUR and the RTN European Program MRTN-CT-2004-503369
as well as (A.A.) by research grants FPA2007-66665, SAB2005-0140 and RFBR 05-02-17477, by Programs RNP 2.1.1.1112; LSS-5538.2006.2 .

\appendix

\section{Einstein-Hilbert lagrangian}
\label{sec:conform_Lagr}

We consider a conformally flat space with $D$ dimensions and write the metric as
\be
  g_{AB}(X) =
    A^2(X) \left[ { \eta_{AB} + h_{AB}(X) } \right]
\ee
where the warp factor $A(X)$ is chosen to depend on all coordinates for convenience. The
inverse metric and the determinant up to quadratic order in the fluctuations are
\ba
g^{AB} &=& A^{-2} \left[ {
    \eta^{AB} - h^{AB} + h^{AC} {h_C}^B + \ldots
  } \right];\\\no
\sqrt{g} &=& A^D \left[
    1 + \frac{1}{2} h
    + \frac{1}{8} h^2 - \frac{1}{4} h^{AB} h_{AB}
    + \ldots
  \right].
\ea
All indexes are raised by the five dimensional flat metric $\eta_{AB}$, so that $h=h_{AB}\eta^{AB}$ for example. To avoid misunderstanding, notice that throughout the text we used a four dimensional representation of the gravitational field in terms of $h_{\mu\nu},v_\mu,S$. Therefore, the 4D trace appearing in the text is actually $h_{\mu\nu}\eta^{\mu\nu}$ and differs from the one in the Appendix because of the $S$ field.

Armed with the above formulas one can prove that the Einstein-Hilbert term $\sqrt{g} R$ can be written, up to quadratic order in the fluctuations, as
\ba
  A^{2-D} \sqrt{g} R &=&
    - \, 2(D-1) \frac{A^{,E}_{,E}}{A}
    - (D-1)(D-4) \frac{A^{,E} A_{,E}}{A^2}
    + 2(D-1) \frac{A_{,EF}}{A} h^{EF}\nonumber \\
  &+& (D-1)(D-4) \frac{A_{,E} A_{,F}}{A^2} h^{EF}
  - \, (D-1) \frac{A^{,E}_{,E}}{A} h
    + 2(D-1) \frac{A_{,E}}{A} {h^{EF}}_{,F}\nonumber \\
  &-& (D-1) \frac{A^{,E}}{A} h_{,E}
    - \frac{1}{2}(D-1)(D-4) \frac{A^{,E} A_{,E}}{A^2} h
    + {h^{EF}}_{,EF} - h^{,E}_{,E} \nonumber \\
  &-& \, 2(D-1) \frac{A_{,EF}}{A} h^{EG} {h_G}^F
    - (D-1)(D-4) \frac{A_{,E} A_{,F}}{A^2} h^{EG} {h_G}^F
    + (D-1) \frac{A^{,G}}{A} h^{EF} h_{EF,G}
    \nonumber \\
  &-& \, 2(D-1) \frac{A^{,G}}{A} h^{EF} h_{GE,F}
    - 2(D-1) \frac{A_{,G}}{A} h^{EG} {h^F}_{E,F}
    + (D-1) \frac{A_{,F}}{A} h^{EF} h_{,E}\nonumber \\
  &+&\, (D-1) \frac{A_{,EF}}{A} h^{EF} h
  +  \frac{1}{2}(D-1)(D-4) \frac{A_{,E} A_{,F}}{A^2} h^{EF} h
    + (D-1) \frac{A_{,E}}{A} {h^{EF}}_{,F} h\nonumber \\
  &-& \,\frac{D-1}{2} \frac{A^{,E}}{A} h_{,E} h
    - \frac{D-1}{4} \frac{A^{,E}_{,E}}{A} h^2
  -  \frac{1}{8}(D-1)(D-4) \frac{A^{,E} A_{,E}}{A^2} h^2\nonumber \\
  &+&\, \frac{D-1}{2} \frac{A^{,G}_{,G}}{A} h^{EF} h_{EF}
    + \frac{1}{4}(D-1)(D-4) \frac{A^{,G} A_{,G}}{A^2} h^{EF} h_{EF}
    \nonumber \\
  &+& \, \frac{3}{4} h_{EF,G} h^{EF,G}
    - \frac{1}{2} h_{EF,G} h^{GF,E}
    + {h^{EF}}_{,F} h_{,E}
    - {h^{EG}}_{,E} {h^F}_{G,F}
    - \frac{1}{4} h_{,E} h^{,E} \nonumber \\
  &+& \, h^{EF} h_{,EF}
    + h^{EF} h_{EF,G}^{,G}
    - 2 {h_E}^F {h^{EG}}_{,GF}
    + \frac{1}{2} {h^{EF}}_{,EF} h
    - \frac{1}{2} h  h^{,E}_{,E}.
\ea

These formulas were first derived in~\cite{Callin}.


\end{document}